# Low temperature dynamics and kinetics of dislocation motion in the high-entropy alloy $Al_{0.5}CoCrCuFeNi$

**V. Natsik, Y. Semerenko, V. Zoryansky**

*B.Verkin Institute for Low Temperature Physics and Engineering of NAS of Ukraine,47 Nauky Ave., Kharkiv, 61103, Ukraine*

semerenko@ilt.kharkov.ua

An analysis of the processes of plastic deformation and acoustic relaxation in a high-entropy alloy $Al_{0.5}CoCrCuFeNi$ was carried out. The following have been established: dominant dislocation defects; types of barriers that prevent the movement of dislocations; mechanisms of thermally activated movement of various elements of dislocation lines through barriers at room and low temperatures. Based on modern dislocation theory, quantitative estimates have been obtained for the most important characteristics of dislocations and their interaction with barriers.

## 1. INTRODUCTION

Multicomponent high-entropy alloys (HEA) discovered at the beginning of the 21st century have a number of improved properties compared to traditional multicomponent alloys. HEA – metal systems of five or more components with a concentration close to equiatomic. Such alloys are characterized by increased values of the entropy of mixing $S$, compared to traditional multicomponent alloys, which explains their name. The explanation for the increased value of entropy in HEA is based on the concept of thermodynamics that the entropy of mixing between soluble components is maximum when these components are in equiatomic concentration and increases with increasing number of components. The change in Gibbs free energy $\Delta G$ when mixing HEA components is determined by the relation: $\Delta G = \Delta H - T\Delta S$, where $\Delta H$ is the change in enthalpy of the system, $T$ is the temperature.

Thus, the contribution of the entropy of mixing during the formation of HEAs reduces the free energy, as a result of which the probability of the formation of substitutional solid solutions with simple crystal lattices in them significantly increases. Indeed, many HEAs have the structure of single-phase solid solutions with fcc or bcc crystal lattices, and such lattices are significantly distorted, since they are formed by atoms of dissimilar elements with different electronic structures and sizes.

Thanks to these features of the atomic structure of HEAs, their properties at room and higher temperatures compare favorably with the properties of traditional metals and alloys: they have a favorable combination of strength and ductility, and high resistance to thermal and mechanical influences. However, during heat treatment and aging of these alloys, phase transformations can be observed, which affects the attractiveness of their practical use.

One of the typical HEA is the $Al_{0.5}CoCrCuFeNi$ with an fcc lattice [1]. Its mechanical and acoustic properties at temperatures $T$<300 K have been studied in detail in [2-7]. In this publication, we will consider the connection between the properties studied in [4-7] and the features of the dynamics and kinetics of elementary dislocation processes in this alloy. We will also discuss the possibility of using the fundamental principles of modern dislocation theory to interpret these properties. We analyse data from two different experimental methods:

- method of resonant mechanical spectroscopy – excitation in samples of cyclic elastic deformation with an amplitude of $\varepsilon_0 \sim 10^{-7}$, caused by short segments of dislocation strings (dislocation relaxers), which oscillate with amplitudes on the order of the lattice parameter [5, 6];
- method of X-ray analysis.

## 2. STUDYED SAMPLES AND METHODS

The production method and characteristics of the alloy are described in [5].

The alloy was studied in 2 structural states: (**I**) - initial cast; (**II**) - after high-temperature annealing in vacuum at 975 °C for 6 hours.

It has been shown [3] that the distribution of elements included in the alloy is non-uniform at the nanoscale. In the microstructure of the alloy, regions are observed in the form of stripes 15-20 nm wide, with significantly different concentrations of various elements; such regions form a three-dimensional irregular lattice in the microstructure. Significant structural distortions are observed near the lattice nodes. The distance between these inhomogeneities is equal to ~ 20 nm. In addition, clusters of several atoms of one of the constituent elements of the alloy are observed. Several adjacent atoms of an element with a relatively large atomic radius create local distortions of the crystal lattice. The characteristic distance between clusters is several nanometers.

The dislocation density in the studied samples was assessed by X-ray analysis.

## 3. EXPERIMENTAL RESULTS

The temperature dependences of internal friction $Q^{-1}_{\exp}(T)$ and dynamic Young's modulus $E_{\exp}(T)$ were studied in [6] using the method of mechanical resonance spectroscopy. The technique for measuring acoustic absorption (internal friction) and dynamic Young's modulus in these experiments is described in [8].

As the temperature decreases, $E_{\exp}(T)$ for state (**I**) increases monotonically and $Q^{-1}_{\exp}(T)$ decreases monotonically, while the temperature dependences $Q^{-1}_{\exp}(T)$ and $E_{\exp}(T)$ do not show any significant features such as relaxation resonances.

The transition to state (**II**) leads to the appearance of a relaxation resonance - an acoustic absorption peak and a corresponding step in the temperature dependence of the dynamic modulus (Fig. 1).

To interpret the acoustic relaxation resonances observed in experiments [6], it is necessary to carry out initial processing of the measurement results of $E_{\exp}(T)$ and $Q^{-1}_{\exp}(T)$: to identify on these temperature dependences the resonant contributions $E_R(T)$ and $Q^{-1}_R(T)$ individual subsystems of relaxers against the background of contributions $E_{BG}(T)$ and $Q^{-1}_{BG}(T)$ other relaxation processes.

In general, the dynamic modulus of elasticity depends on both temperature $T$ and frequency $\omega$. However, our measurements were performed at a fixed value of the sample oscillation frequency $\omega_r = 2\pi f_r$. It has been established [8] that the dependence $E_{\exp}(T,\omega)$ recorded in experiments can be divided into resonant $E_R(T,\omega)$ and background $E_{BG}(T,\omega)$ components:

$$E_{\exp}(T,\omega) = E_0(\omega) - E_{BG}(T,\omega) - E_R(T,\omega) \tag{1}$$

where $E_0(\omega)$ is the limit value of the module at $T \to 0$. For most crystalline materials, the value $E_0$ is close to the value of the static modulus of elasticity of a dislocation-free crystal.

According to [9], for many crystalline materials the background component $E_{BG}(T,\omega)$ in the region of low temperatures $T < 300$ K and frequencies $\omega \le 10^7$ s$^{-1}$ is determined primarily by the interaction of elastic vibrations with thermal phonons. For the Einstein model of the phonon spectrum with a characteristic temperature $\Theta_E$, the softening of the elastic modulus by phonons is described by the formula

$$\frac{E_{BG}(T,\omega)}{E_0(\omega)} = \beta \cdot T \exp\left(-\frac{T_\eta}{T}\right), \tag{2}$$

where the coefficient $\beta$ depends on the material under study and the vibration mode under study, and the characteristic temperature $T_\eta$ in most cases for materials with simple phonon spectra is of the order of the Einstein $\Theta_E$ or Debye $\Theta_D$ temperatures: $T_\eta \le \Theta_E = h\nu_E \approx \frac{3}{4}\Theta_D$ ($h$ - Planck constant), since in real crystals the frequencies of acoustic phonons are lower than the Einstein frequency $\nu_E$ [10]. In the simplest case $\Theta_E = \frac{3}{4}\Theta_D$.

In the temperature range T ≤ 10 K, along with the phonon contribution (2), one can also distinguish a relatively weak electronic contribution in $E_{BG}(T,\omega)$ [8], but it does not play a significant role in the analysis of dynamic elastic moduli in the temperature range $T < \Theta_D$.

Then: $$E_{\exp}(T,\omega) = E_0(\omega)\cdot\eta(T) - E_R(T,\omega), \quad \eta(T) = 1 - \beta T \exp(-T_\eta/T) \qquad (3)$$

It has been shown [5] that (3) well describes the $E_{\exp}(T,\omega)$ of alloy under study in the absence of relaxation resonances, when $E_R(T,\omega) \equiv 0$ (state **I**). Its use for approximating the results obtained when studying state (**II**) is illustrated in Fig. 1a (solid line), and the corresponding parameter values are given in Table 1. When transitioning from state (**I**) to state (**II**), the parameter $E_0$ increases, but the characteristic temperature $T_\eta$ and coefficient $\beta$ remain unchanged.

The temperature dependence of the resonant component of the Young's modulus for state (**II**) is shown in Fig. 1a. Its graph has the shape of a step with a height characteristic of relaxation resonances $E_{R0} = E_R(T \gg T_p, \omega)$ (Table 1).

$$E_R(T,\omega) = E_0(\omega)\cdot\eta(T) - E_{\exp}(T,\omega) \qquad (4)$$

Table 1

Parameters of background $E_{BG}(T,\omega)$ and $Q_{BG}^{-1}(T)$ for HEA $Al_{0.5}CoCrCuFeNi$ in state (**II**) [6]

| $\omega = 2\pi f_r$ | $E_0$ | $\beta$ | $T_\eta$ | $E_{R0}$ | $A_1$ | $A_2$ | $U_{BG}$ |
|---|---|---|---|---|---|---|---|
| $3.34\cdot10^3$ s$^{-1}$ | 236 GPa | $3.5\cdot10^{-4}$ K$^{-1}$ | 160 K | 0.33 GPa | $6\cdot10^{-5}$ | 0.3 | 0.16 eV |

The experimentally observed dependence $Q_{\exp}^{-1}(T,\omega)$ consists of the sum of resonant $Q_R^{-1}(T,\omega)$ and background $Q_{BG}^{-1}(T,\omega)$ absorption. The interpretation of the internal friction peaks recorded in the experiment comes down to a comparison with the theory of the difference value

$$Q_R^{-1}(T,\omega) = Q_{\exp}^{-1}(T,\omega) - Q_{BG}^{-1}(T,\omega) \qquad (5)$$

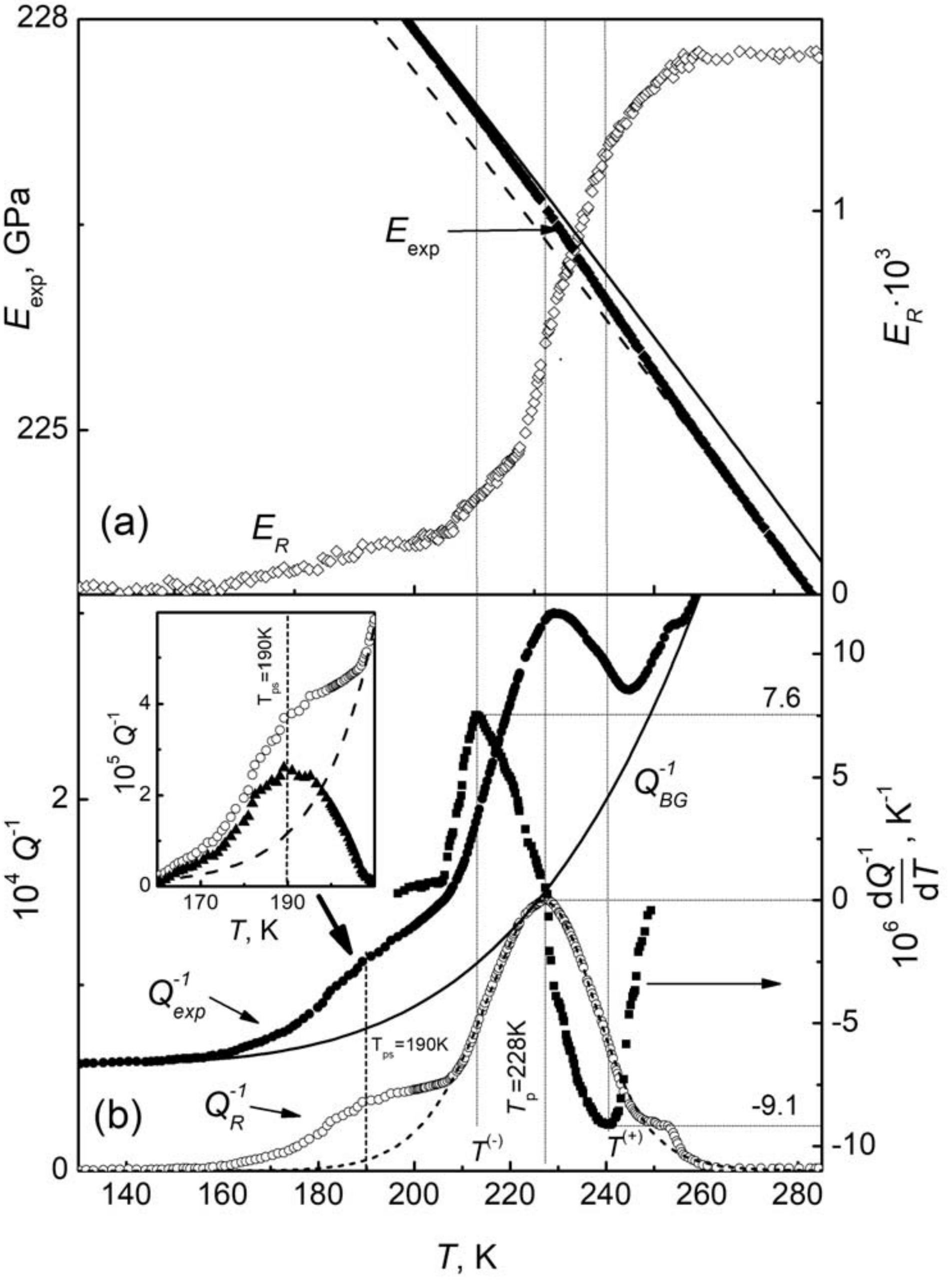

Fig. 1 Relaxation resonances in the $Al_{0.5}CoCrCuFeNi$ alloy in the structural state (**II**) [6]: a) temperature dependence of dynamic Young's modulus: ♦ - $E_{\exp}(T)$, ◊ - $E_R(T)$; b) temperature dependence of internal friction: ● - $Q^{-1}_{\exp}(T)$, ○ – $Q^{-1}_R(T)$. Solid lines show the background of the dynamic modulus $E_0 - E_{BG}(T)$ and the absorption background $Q^{-1}_{BG}(T)$: these graphs are built on the basis of formulas (2) and (6) with the parameter values given in Table. 1. In Fig. 2b also shows the results of numerical differentiation of resonant absorption near temperature $T_p$: ■ – $\frac{\partial}{\partial T} Q^{-1}_R(T,\omega)$. The dotted lines are drawn through the inflection points $T^{(-)}$, $T^{(+)}$ and the top of the peak $T_p$, and the dashed line shows the theoretical dependence calculated according to formula (17) with the parameter values given in Table 2.

The inset in Fig.1b shows: ○ – resonant component of internal friction $Q^{-1}_R(T)$; the dashed line shows the approximation $\langle Q^{-1}_R \rangle$ for low temperature peak slope $T_p = 228$ K; ▲ – absorption peak satellite $T_{ps} = 190$ K.

We will consider [8] the $Q_{BG}^{-1}(T)$ to be partially caused by thermally activated dislocation relaxation with an activation energy $U_{BG}$, which differs significantly from the activation energy of the studied resonance absorption. For description $Q_{BG}^{-1}(T)$ we use the relation [11]:

$$Q_{BG}^{-1}(T) = A_1 + A_2 \exp\left(-\frac{U_{BG}}{k_B T}\right). \tag{6}$$

$A_1$, $A_2$ and $U_{BG}$ are fitting parameters. The coefficient $A_1$ characterizes the contributions to the absorption of the phonon, electronic and magnetic subsystems of the metal, which weakly depend on temperature near resonance.

Relationship (6) well describes the temperature dependence of the acoustic absorption background for state (**II**) - solid line in Fig. 1b ($A_1$, $A_2$ and $U_{BG}$ are given in Table 1).

The temperature dependence $Q_R^{-1}(T,\omega)$ of the alloy under study in state (**II**), after subtracting the background $Q_{BG}^{-1}(T)$, is shown in Fig. 1b. Analysis of the temperature dependence of the derivative $\frac{\partial}{\partial T} Q_R^{-1}(T,\omega)$ makes it possible to clarify the peak temperature $T_p$, obtain the values of the coordinates of the inflection points $T^{(-)}$ and $T^{(+)}$ on the graph $Q_R^{-1}(T,\omega)$ (see Fig. 1b), and also estimate the value of the ratio

$$K = \frac{\max \frac{\partial}{\partial T} Q_R^{-1}(T,\omega)}{\left|\min \frac{\partial}{\partial T} Q_R^{-1}(T,\omega)\right|} \tag{7}$$

Registration in experiments of the characteristics of acoustic relaxation resonance $T_p$, $T^{(-)}$, $T^{(+)}$, $\max Q_R^{-1}$, $E_{R0}$ and $K$ (see Table 2) allows us to formulate a microscopic model of the relaxer and obtain estimates for its parameters [8, 12].

However, the resonant component of acoustic relaxation in state (**II**) is not limited only to the contributions of relaxers responsible for the appearance of the peak $T_p = 228$ K. Fig. 1b shows that the resonant component $Q_R^{-1}(T,\omega)$ contains another relaxation resonance, localized on the left slope of the peak $T_p = 228$ K - a satellite $T_{ps} = 190$ K. Its characteristics were obtained further after a detailed analysis of the main peak.

Table 2

Characteristics of the main absorption peak (a) and its satellite (b) in the state (**II**)

(a) [6]

| $T_p$ | $T^{(-)}$ | $T^{(+)}$ | $\max Q^{-1}$ | $\max \frac{\partial}{\partial T}\bar{Q}^{-1}$ | $\min \frac{\partial}{\partial T}\bar{Q}^{-1}$ | $K$ |
|---|---|---|---|---|---|---|
| 228 K | 213 K | 242 K | $1.5 \cdot 10^{-4}$ | $7.6 \cdot 10^{-6}$ K$^{-1}$ | $-9.1 \cdot 10^{-6}$ K$^{-1}$ | 0.83 |

(b) [33]

| $T_{ps}$ | $T_s^{(-)}$ | $T_s^{(+)}$ | $\max Q_s^{-1}$ | $\max \frac{\partial}{\partial T}\bar{Q}_s^{-1}$ | $\min \frac{\partial}{\partial T}\bar{Q}_s^{-1}$ | $K_s$ |
|---|---|---|---|---|---|---|
| 190 K | 182.6 K | 202.0 K | $2.64 \cdot 10^{-5}$ | $3.1 \cdot 10^{-6}$ K$^{-1}$ | $-2.5 \cdot 10^{-6}$ K$^{-1}$ | 1.24 |

## 4. LOW TEMPERATURE DISLOCATION PROCESSES IN HEA $Al_{0.5}CoCrCuFeNi$

The alloy under study in states (**I**) and (**II**) has the morphology of a polycrystalline with an fcc lattice. Therefore, it is natural to assume that the features of the initial stage of plastic deformation and acoustic relaxation resonances in this alloy are due to the dynamics and kinetics of dislocation processes similar to low-temperature dislocation processes in polycrystals of monatomic materials with an fcc structure [8, 13, 15-21]. Differences in the atomic structure and morphology of grain boundaries in HEAs and simple metals must be taken into account only when interpreting experimental results obtained under conditions of sufficiently high temperatures, when their acoustic and mechanical properties are significantly influenced by the processes of thermally activated diffusion transformation of grain boundaries of polycrystals, in particular, absorption and generation of dislocations that determine plastic deformation.

The interpretation of the laws of dislocation plasticity and the influence of dislocations on the acoustic properties of metals is based on the idea of the presence of easy slip planes in their structure and the specificity of the nucleation and movement of dislocations in these planes [22, 23]. Currently, experimental methods for observing such processes have been developed, as well as theoretical descriptions of the results of these experiments based on concepts of the dynamics and kinetics of thermally activated and quantum motion of dislocations in easy slip planes through various barriers, taking into account their inhibition not only by barriers, but also by quasiparticles - conductivity electrons and phonons [6, 8, 12].

## 4.1 Models of dislocation relaxers

Let us use the algorithm [8, 12] for analyzing mechanical spectroscopy data, which allows us to establish the microscopic mechanism of relaxation resonances and obtain estimates for the parameters of elementary relaxers based on an analysis of the temperature dependences $Q^{-1}_{\exp}(T)$ and $E_{\exp}(T)$ obtained in experiments at one fixed value of the sample oscillation frequency $\omega_r = 2\pi f_r$.

Since the bulk of the samples of the studied alloy, both in state (**I**) and state (**II**), consists of a material with an fcc lattice, it is natural to assume that the observed acoustic relaxation resonances in this alloy are determined by mechanisms that are similar to resonances in metals with an fcc structure. Such resonances are interpreted as a consequence of the interaction of elastic vibrations of the sample with a system of dislocation relaxers, since they are observed only after preliminary plastic deformation of the samples. On the temperature dependence of internal friction, they correspond to Bordoni peaks in the range 20 K < $T$ < 100 K [15-18] and Hasiguti peaks in the range 100 K < $T$ < 200 K [24]. The positions of these peaks $T_p(\omega)$ on the temperature axis depend on the vibration frequency $\omega$ and differ for different metals.

In fcc crystals at the initial stage of deformation, dislocation plastic sliding predominantly occurs along close-packed <110> directions and {111} planes (Fig. 2). In the {111}<110> slip system, there are two types of rectilinear dislocations with the Burgers vector $\mathbf{b}$ whose lines $D^{qe}$ and $D^{s}$ are oriented along the directions $ox$ of dense packing. But for $D^{qe}$ the vector $\mathbf{b}^{qe}$ has a quasi-edge (close to the edge) orientation, and for $D^{s}$ the vector $\mathbf{b}^{s}$ has a purely s screw orientation: the angle between the vector $\mathbf{b}^{qe}$ and the unit $\boldsymbol{\phi}$ of the dislocation line has the value $\alpha = 60^{o}$, and the angle between $\mathbf{b}^{s}$ and $\boldsymbol{\phi}$ is equal to zero, while $\left|\mathbf{b}^{qe}\right| = \left|\mathbf{b}^{s}\right| = \left|\mathbf{b}\right| = \frac{\sqrt{3}}{2} = a_0$, $a_0$ is the distance between the nearest nodes.

Moving lines in the transverse direction The movements of rectilinear dislocation lines in the transverse direction $oz$ are controlled by the first kind Peierls lattice potential relief with period $a_{p1}$, but the height of the barriers of this relief and the corresponding Peierls critical stress values $\tau^{s}_{p1}$ and $\tau^{qe}_{p1}$ for pure screw dislocations are significantly greater than for quasi-edge ($\tau^{s}_{p1} \gg \tau^{qe}_{p1}$).

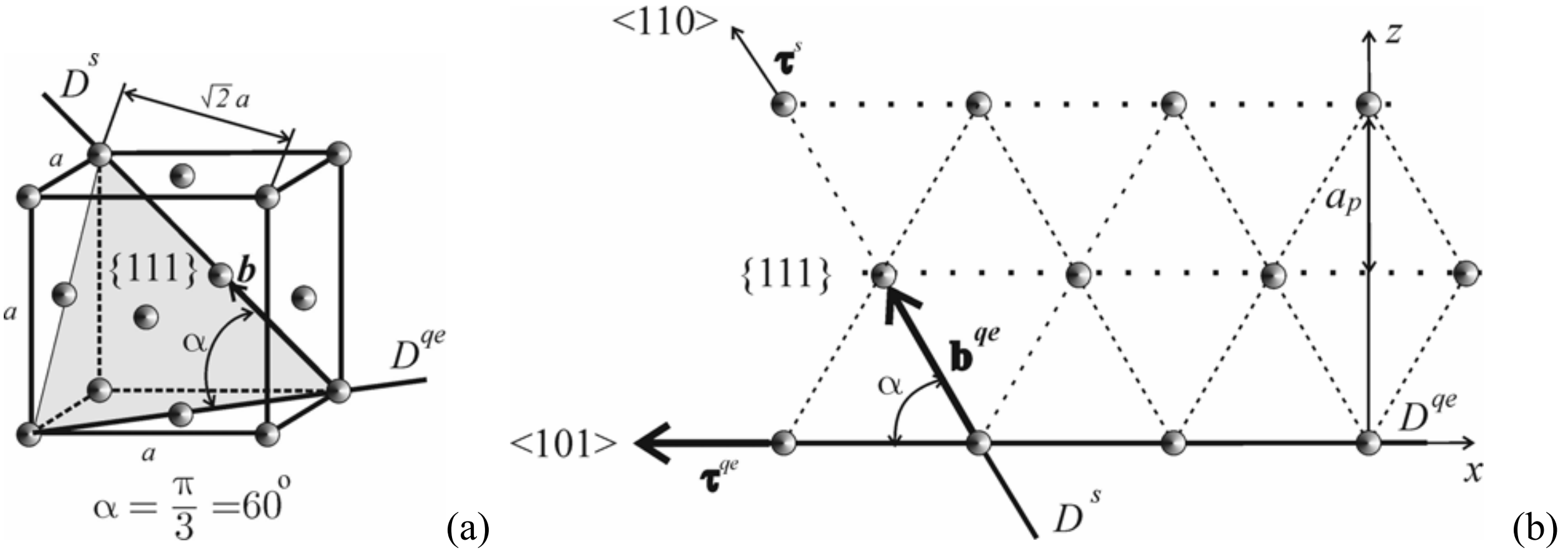

Fig. 2. {111}<110> slip system and straight dislocations in an fcc crystal:
(a) – unit cell; (b) – one of the sliding planes {111} [6].

● , $a$- nodes and fcc lattice parameter, $a \simeq 0.36$ nm [6]; $D^{qe}$, $D^{s}$ – lines of quasi-edge and screw dislocations with a common Burgers vector $\mathbf{b}^{qe} = \mathbf{b}^{s} = \mathbf{b} = b\langle 0\bar{1}1\rangle$ with length $b^{s} = b^{qe} = b = a_0 = \dfrac{a}{\sqrt{2}}$, $b = 0.254$ nm [13]; $a_p = \dfrac{\sqrt{3}}{2}a_0 = \sqrt{\dfrac{3}{8}}a$ - Peierls relief period in the direction of easy sliding; $a_0$ - distance between adjacent nodes in the direction of easy sliding.

When such dislocations nucleate and move in real materials with structural defects on the surface and in the bulk, the formation of only straight configurations of dislocation lines is unlikely [25]. At the early stages of plastic deformation, a set of curved dislocation segments is formed (see Fig. 5) with ***ABC*** configurations between the attachment points, which consist of shorter fragments ***AB*** and ***BC*** with significantly different crystal geometric and dynamic properties. Straight-line segments of the dislocation line ***BC*** with a length $L$ are oriented along the directions of dense packing and are located in the valleys of the Peierls relief, and fragments ***AB*** with a length $\tilde{L}$ consist of chains of kinks connecting short straight-line segments of dislocation lines in neighboring valleys of the relief. The self-energy of an individual kink also has a periodic component if its center $x_k$ moves along the direction of close packing parallel to the axis $ox$: this is called the secondary Peierls relief. The period of this relief is equal to the minimum distance between nodes $a_{p2} = a_0$, and the height of the barriers and the values of critical stresses $\tau^{s}_{p2}$ and $\tau^{qe}_{p2}$ are different for $D^{qe}$ and $D^{s}$ dislocations ($\tau^{s}_{p2} \ll \tau^{qe}_{p2}$).The ***ABC*** segment has the properties of a two-mode dislocation relaxer, which consists of fragments $L$ and $\tilde{L}$. The significant difference in the crystal geometric and energy characteristics of the fragments also determines the difference in the dynamic and relaxation properties of these components $L$ and $\tilde{L}$ of the relaxer during its interaction with elastic vibrations of the sample.

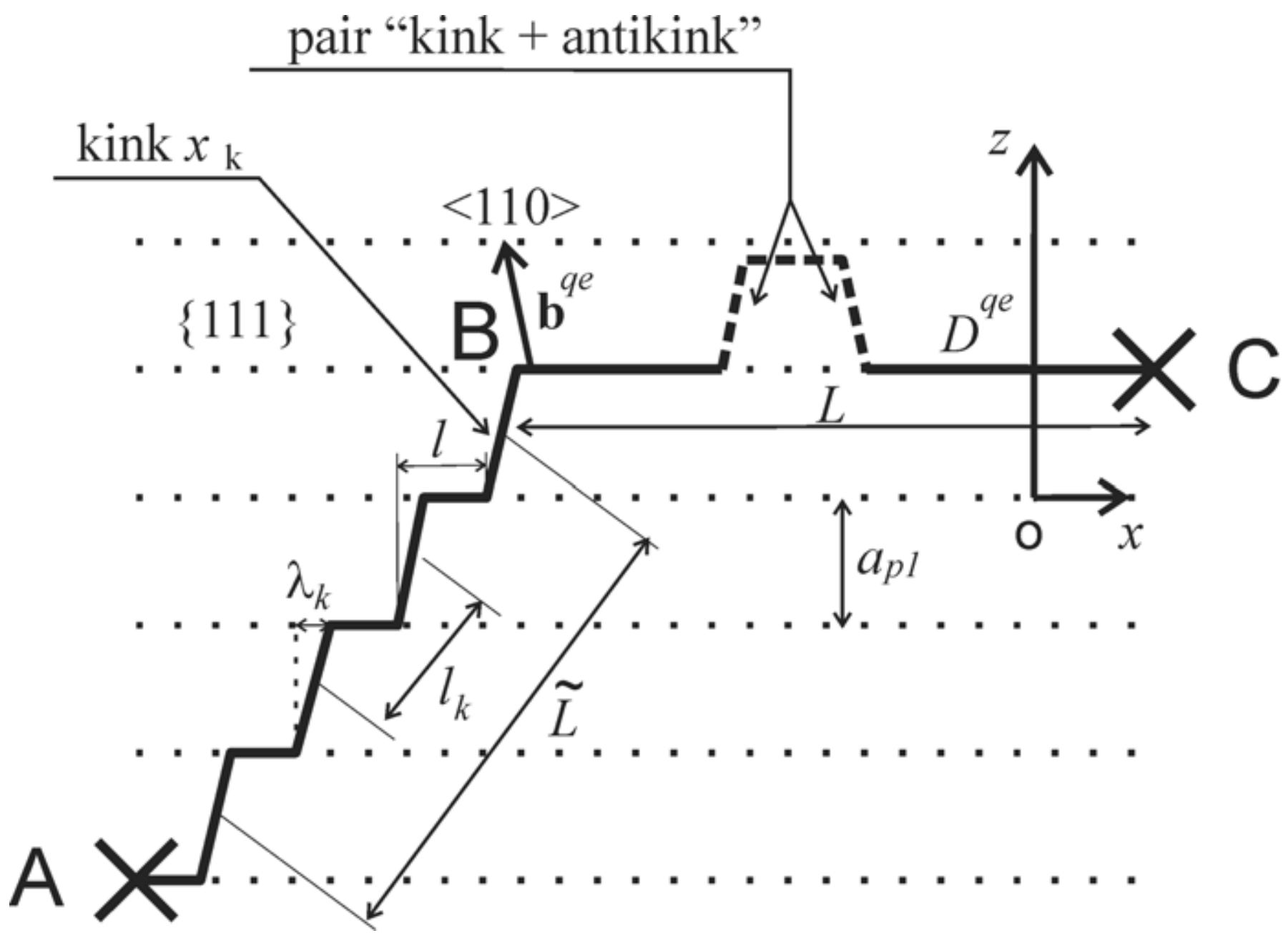

Fig. 3. Configurations of dislocation lines in the {111}<110> slip system in an fcc crystal [6]: ***ABC*** – curved segment of a quasi-edge dislocation $D^{qe}$ with a Burgers vector $\mathbf{b}^{qe}$, the dotted line indicates the close packing directions; $a_{p1}$ - period of the first kind Peierls relief in the direction of the axis $oz$; $a_0=b$ - period of secondary Peierls relief; $L$ - length of straight segment ***BC*** in the relief valley; $\tilde{L}$ - length of the chain of ***AB*** kinks between relief valleys; $x_k$ - coordinate of a separate kink along the axis $ox$; $\lambda_k$ - kink width; $l$ – distance between the centers of neighbouring kinks.

It is known that the elementary relaxers for the Bordoni and Hasiguti peaks are fragments of dislocation lines in easy slip systems $\{111\}\langle 110\rangle$ of fcc metals, which are excited by elastic vibrations [21, 24]. In the proposed model of a two-mode dislocation relaxer, the role of such fragments is played by its components ***BC*** (length $L$) and ***AB*** (length $\tilde{L}$). It is assumed that their ends ***A***, ***B*** and ***C*** are fixed by local defects in the lattice structure, which prevent the movement of the dislocation line in the slip plane.

Bordoni peaks were interpreted by Seeger [19-21] as a result of thermally activated nucleation of paired kinks on straight segments of dislocation lines ***L*** located in the valleys of the Peierls relief (Fig. 4a).

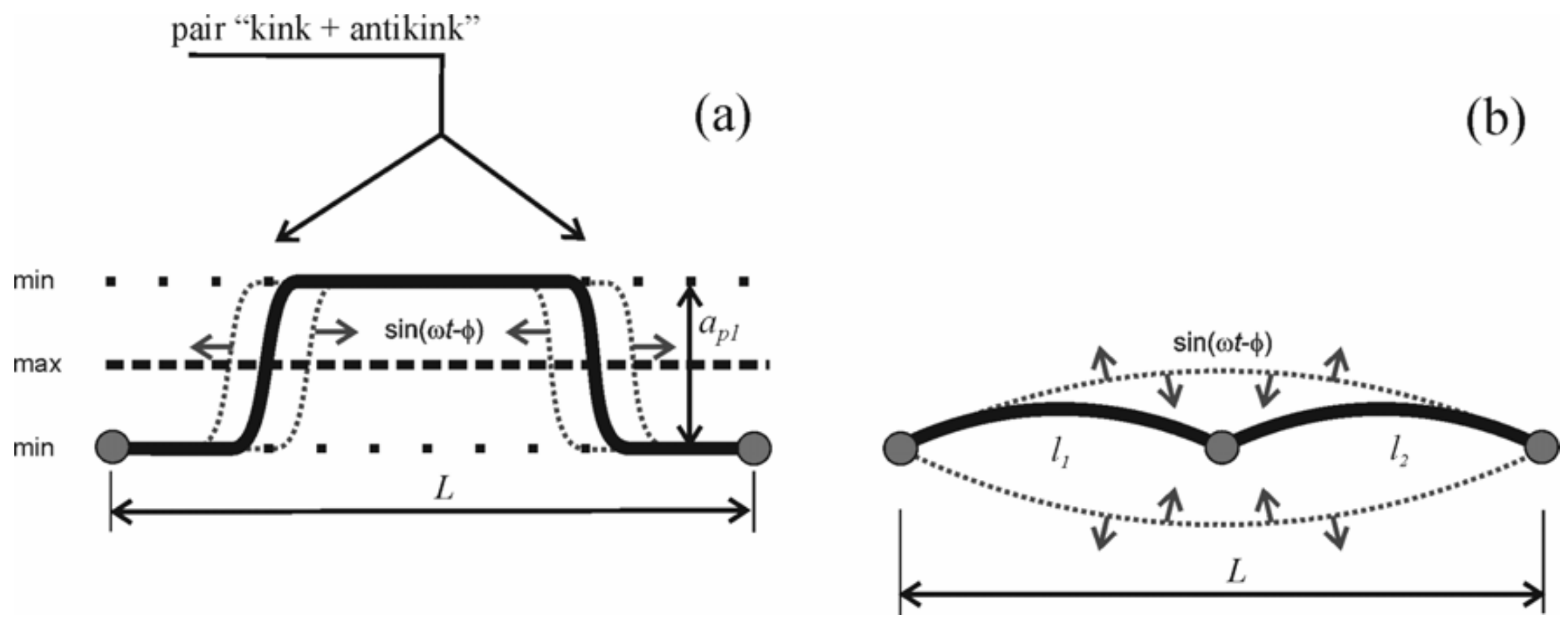

Fig. 4 Schematic representation of an elementary relaxer [6]: (a) – Seeger relaxation; (b) – relaxation Koiwa and Hasiguti; ● – local defects on the dislocation string; the symbol *L* denotes the length of the dislocation segment, the activation of which determines the elementary contribution of the dislocation to the acoustic resonance or the rate of plastic deformation.

For Hasiguti peaks, Koiwa and Hasiguti [24] considered the thermally activated detachment of a dislocation line segment $L = l_1 + l_2$ from an individual point defect (impurity atom, vacancy, radiation defect, etc.) as an elementary relaxation process (Fig. 4b). It has been established that vibrations of a chain of geometric kinks on a segment of a dislocation line with fixed ends are described by the equation of vibrations of a segment of a dislocation string [23, 25].

## 4.2 Thermal activation and statistical analysis of the dislocation contribution to acoustic relaxation

A separate elementary statistically independent process of thermally activated excitation of a dislocation relaxer is characterized by a relaxation time, which depends on temperature according to the Arrhenius law:

$$\theta(T) = \theta_0 \exp\left(\frac{U_0}{k_B T}\right) \tag{9}$$

$U_0$ (activation energy) and $\theta_0$ (effective period of attempts) are determined by the crystal geometric and energy characteristics of a particular relaxer in a particular crystal. The $\theta_0$ has a weak power-law dependence on temperature, which can be neglected against the background of the exponential dependence of the second factor at low temperatures $kT \ll U_0$.

The temperature-frequency dependence of internal friction, caused by a system of relaxers with the same values of parameters $U_0$ and $\theta_0$, is described by the formula

$$Q_R^{-1}(T,\omega) = C_r \Delta_0 F(\omega\theta) \tag{10}$$

Here $\Delta_0$ and $C_r$ are, respectively, the effective specific contribution of an individual relaxer and their concentration, and $F(\omega\theta)$ is a positive definite function with a sharp maximum at $\omega\theta \simeq 1$. The form of the function $F(\omega\theta)$ and the position of its maximum on the temperature axis $T_p(\omega)$ depend on the nature of the relaxers. Most of them belong to the class of Debye relaxers, for which thermal activation occurs simultaneously in the forward and reverse directions with respect to the exciting voltage; the Seeger relaxation meets this criterion and is a special case of the Debye relaxation. But in the Koiwa-Hasiguti process, thermal activation stimulates the excitation of relaxers only in the forward direction, and their return to the initial unexcited state occurs under the action of the linear tension force of the dislocation segments. This results in two different expressions for the function $F(\omega\theta)$ [12, 24]:

$$F^D(x) = \frac{x}{1+x^2}, \qquad F^{K-H}(x) = \frac{x^2}{2(1+x^2)}\left[1-\exp\left(-\frac{2\pi}{x}\right)\right], \quad x = \omega\theta \tag{11}$$

For relaxation processes of both types, the function $F^{D,K-H}(x)$ vanishes at $x \to 0$ and $x \to \infty$, but has a sharp maximum at $x_m^D = 1$ and $x_m^{K-H} = 2.67$, respectively. If the value of the oscillation frequency $\omega$ is fixed, then for a system of identical relaxers with parameters $U_0$ and $\theta_0$ internal friction $Q_R^{-1}(T,\omega)$ on the temperature axis has a sharp maximum (peak) at temperature $T_p^{D,K-H}(\omega)$:

$$T_p^{D,K-H}(\omega) = \frac{U_0}{k_B \alpha^{D,K-H}}; \; \alpha_D = -\ln \omega\theta_0, \quad \alpha_{K-H} = -\ln\frac{\omega\theta_0}{2.67} \approx 1 - \ln\omega\theta_0 \tag{12}$$

Parameter estimates obtained $U_0$, $\theta_0$, $C_r\Delta_0$:

- <u>for Seeger relaxers</u>

$$U_0 \sim 0.1 \text{ eV}, \; \theta_0 \sim 10^{-11} \text{ s and } C_r\Delta_0 \sim 10^{-1}\rho_L L^3, \tag{13}$$

where $L$ is the length of a straight dislocation segment in the Peierls relief valley, and $\rho_L$ is the number of such segments per unit volume;

- <u>for Koiwa-Hasiguti relaxers</u>

$$U_0 \sim (0.3 \div 0.5) \text{ eV}, \; \theta_0 \sim 10^{-13} \text{ s and } C_r\Delta_0 \sim 10^{-1}\rho_L L^3, \tag{14}$$

where $L$ and $\rho_L$ are, respectively, the length and volume density of dislocation segments breaking away from an individual point defect.

The action of the relaxation process is accompanied by a decrease in the dynamic modulus of elasticity by the amount $E_R(T,\omega)=E_R(\omega\theta)$. In this case $E_R(\omega\theta\to\infty)=0$, and a decrease of $\omega\theta$ leads to a monotonic increase $E_R(T,\omega)$ to $E_{R0}=E_R(\omega\theta\to 0)$.

Let us note two differences between the Debye and Koiwa-Hasiguti processes:

- in [24] it is shown that these processes correspond to different values of the ratio of the peak height $\max Q_R^{-1}(\omega\theta)$ to the step height $E_{R0}=E_R(\omega\theta\to 0)-E_R(\omega\theta\to\infty)$ on the temperature-frequency dependences of the contributions of these relaxation processes to the internal friction and dynamic elasticity of materials:

$$\left[\frac{E_0}{E_{R0}}\cdot\max Q_R^{-1}\right]^D=0.5\,,\quad \left[\frac{E_0}{E_{R0}}\cdot\max Q_R^{-1}\right]^{K-H}\approx 0.13 \qquad (15)$$

- in [12] it was established that the parameter $K$ determined by relation (7) does not depend on temperature and $U_0$, and its frequency dependence is described by monotonic functions $K=K_{D,K-H}(\omega\theta_0)$, while:

$$K_D(\omega\theta_0)>1.2\,,\quad K_{K-H}(\omega\theta_0)<1.2\,,\quad \omega\theta_0<10^{-3} \qquad (16)$$

Relations (9)-(16) describe the acoustic relaxation resonance in an ideal crystal, caused by a system of similar dislocation relaxers with the same values of the parameters of an individual relaxer $U_0$, $\theta_0$ and $\Delta_0$. In a real HEAs, there is a complex system of random structural defects or inhomogeneities and the internal stress fields they create. Therefore, the parameters of the same type relaxers $U_0$, $\theta_0$ and $\Delta_0$ acquire random additives in different areas of the sample, which leads to a statistical broadening of peaks and steps in the dependence graphs $Q_R^{-1}(T,\omega)$ and $E_R(T,\omega)$; and also to a shift in the temperature $T_p(\omega)$ of their localization. In HEAs, random inhomogeneities are associated not only with the chaotic distribution of defects, but also with distortions of unit cells in the crystal lattice by random configurations of alloy components and differences in their atomic radii [26].

Following [8, 12], we will consider $U_0$, $\theta_0$ and $\Delta_0$ as random variables with their corresponding distribution patterns. But at low temperatures $k_BT\ll U_0$ it is enough to take into account only statistical deviations of the activation energy $U$ from $U_0$ and with exponential accuracy we can neglect the scatter of parameters $\theta_0$ and $\Delta_0$. In this case, the contribution of relaxers to internal friction $\left\langle Q_R^{-1}(T,\omega)\right\rangle$ and dynamic modulus of elasticity $\left\langle E_R(T,\omega)\right\rangle$ will be determined by averaging the initial formulas for the dependencies $Q_R^{-1}(T,\omega)$ and $E_R(T,\omega)$ with a quasi-Gaussian distribution function $P(U;U_0,D)$ for the activation energy, in which the role of parameters is played by activation energy $U_0$ and its dispersion characteristics $D$ [6, 8, 12].

After averaging, relation (10) takes the form:

$$\langle Q_R^{-1}\rangle = Q_R^{-1}(T,\omega;\theta_0,U_0,D) = C_r\Delta_0\int_0^\infty dUP(U;U_0,D)F^{D,K-H}(\omega\theta), \quad (17)$$

The statistical spread of activation energy does not affect the form of functions $K_{D,K-H}(\omega\theta_0)$ and preserves (16), which allows us to establish a physical model of a dislocation relaxer.

The statistical spread of activation energy leads to a decrease in the height of the internal friction peak $\max\langle Q_R^{-1}\rangle < \max Q_R^{-1}$ and an increase in its width; and also increases the width of the step on the temperature dependence of the module $\langle E_R(T,\omega)\rangle$, but maintains its height $E_{R0}$. Therefore, at a small value of the parameter $D \ll U_0$, relations (15) are approximately preserved and can be used to select a relaxer model.

## 4.3 Dislocation mechanism of the internal friction peak $T_p = 228$ K (analogue of the Hasiguti peak)

Comparing the resonance parameters (Table 2) near temperature $T_p = 228$ K with relations (15) and (16), we come to the conclusion that it corresponds to the Koiwa-Hasiguti process, i.e. thermally activated separation of dislocation segments from point defects:

$$K = 0.83 < 1.2, \qquad \frac{E_0}{E_{R0}}\cdot\max Q_R^{-1} \approx 0.1$$

In the studied alloy, the role of point defects can be played by nanoclusters of several atoms of one of the chemical elements of the alloy [3] or traditional point defects of the crystal structure - vacancies and interstitial atoms. Therefore, for a theoretical description of resonance, we use (17), assuming that $F(\omega\theta) = F^{K-H}(\omega\theta)$, and in the future we will omit the index “K-H”. In [6], relations were obtained that allow one to estimate the values of the parameters $U_0$, $\theta_0$, $\Delta_0$ and $D$ if we use the experimentally recorded values of the resonance characteristics $T_p$, $T^{(-)}$, $T^{(+)}$, $\max Q_R^{-1}$, $E_{R0}$ and $K$ (see Table 2):

$$(\alpha-1)^{1,43} = \frac{10}{K-0,7}, \quad \omega\theta_0 = \exp(1-\alpha); \quad C_r\Delta_0 = 2.5\max Q_R^{-1}\cdot\exp\left(\frac{7\sqrt{2}\alpha D}{56\sqrt{2}D+20U_0}\right) \quad (18)$$

$$U_0 = 8k_B\alpha^2\cdot\left(2T_p - T^{(-)} - T^{(+)}\right), \qquad D = \frac{20k_B\alpha}{11\sqrt{2}}\cdot\left(9T^{(+)} + 10T^{(-)} - 19T_p\right) \quad (19)$$

Substituting into (18) and (19) the values of the resonance characteristics from Table 2 we obtain estimates for the relaxer parameters $U_0$, $\Delta_0$ and $\theta_0$ (Table 3), which allows us to calculate the theoretical profile $\langle Q_R^{-1} \rangle$ of the internal friction peak (solid line in Fig. 7).

The obtained values of activation energy $U_0 \approx 0.4$ eV and effective oscillation period $\theta_0 \approx 10^{-13}$ s are typical for Koiwa-Hasiguti relaxers in simple metals with an fcc structure. A small value $D \sim 10^{-2} U_0$ indicates the absence of significant structural distortions in state (**II**), which is due to long-term annealing, which was used to form this structural state.

Table 3

Parameters of dislocation relaxers for the peak of internal friction $T_p$ and its satellite $T_{ps}$

| $\theta_0$ | $U_0$ | $D$ | $C_r\Delta_0$ | $\theta_0^s$ | $U_0^s$ | $D^s$ | $C_r^s\Delta_0^s$ |
|---|---|---|---|---|---|---|---|
| $2\cdot10^{-13}$ s | 0.43 eV | 0.01 eV | $4\cdot10^{-4}$ | $4\cdot10^{-11}$ s | 0.07eV | 0.01 eV | $1\cdot10^{-4}$ |

Using (14) and assuming $C_r \approx \rho_L$ we obtain an estimate $\Delta_0 \sim 10^{-1} L^3$ for the contribution of one relaxer to internal friction and decreasing of the elastic modulus. Let us assume that the role of local centers of pinning of dislocation segments is played by small atomic clusters with a distance between them of the order of several nanometers: they are recorded in state (**II**) by electron microscopy methods [3]. Then, when estimating the length $L$, we can take twice the distance between clusters $L \sim 10$ nm, and given in Table 2 corresponds to the volume density of relaxers $C_r = \rho_L \sim 4 \cdot 10^{21}$ m$^{-3}$. For the dislocation density $\Lambda_d = l\rho_L$ (total length of dislocation segments per unit volume), which effectively interact with elastic vibrations of the sample, we obtain the estimate $\Lambda_d \sim 4 \cdot 10^{13}$ m$^{-2}$. This theoretical estimate correlates with our data obtained by X-ray diffraction analysis.

## 4.4 Dislocation mechanism of internal friction peak $T_{ps} = 190$ K (analogue of Bordoni peak)

A satellite peak with a maximum at $T_{ps} = 190$ K is observed on the low-temperature branch of the main peak $T_p = 228$ K. To isolate it and subsequent analysis, it is necessary to subtract the contribution of theoretical dependence $\langle Q_R^{-1} \rangle$ from the experimentally observed relaxation $Q_{\exp}^{-1}(T)$ at the values of the relaxator parameters $U_0$, $\theta_0$ and $\Delta_0$ corresponding to the main peak (Table 3). The results of this procedure are shown in the inset in Fig. 1b.

Using the statistical and thermal activation analysis described above, we come to the conclusion that the peak $T_{ps} = 190$ K is due to Seeger relaxation, and the estimates for its parameters are given in Table 3.

We will carry out a theoretical description of the movement of a dislocation line in a first kind Peierls relief in the case of a sinusoidal relief, when the linear energy density $W(z_d)$ of an element of a dislocation line has the form:

$$W(z_d) = \Gamma + W_0 \sin^2 \frac{\pi z_d}{a_p} - b\sigma z_d, \quad W_0 = \frac{ba_p}{\pi}\tau_{p1} \tag{20}$$

here $\Gamma$ is the energy per unit length of a dislocation in the continuum approximation (linear tension); $W_0$ and $\tau_{p1}$ - magnitude of barriers and critical stress for Peierls relief of the first type; $\tau = \tau_{xz}$ - shear stress component in the slip plane; $z_d$ - coordinate of the dislocation line element along the axis $oz$ (Fig. 3).

The equation of motion of a string with linear mass density $M$ in the potential (20) has soliton solutions in the form of kinks, and their characteristics are related to the parameters of the potential and dislocation by the formulas [8]:

$$\lambda_k = a_p\sqrt{\frac{2\Gamma}{\pi b a_p \tau_{p1}}}, \quad m_k = \frac{2a_p M}{\pi}\sqrt{\frac{2ba_p\tau_{p1}}{\pi\Gamma}}, \quad \varepsilon_k = m_k c_t^2, \quad \theta_{p1} = \frac{\pi\lambda_k}{c_t}, \quad c_t = \sqrt{\frac{\Gamma}{M}}. \tag{21}$$

Here $\lambda_k$, $m_k$, $\varepsilon_k$ are the width, mass and energy of the kink, respectively, $\theta_{p1}$ is the period of natural oscillations of a rectilinear segment in the relief valley, $c_t$ is the characteristic value of the speed of transverse sound vibrations in the crystal.

The average time of thermally activated nucleation of kink–antikink pairs on a straight segment of a dislocation line (Fig. 4a) is described by the Arrhenius law (9) with activation energy $U \approx 2\varepsilon_k$ and attempt period $\theta_0 \approx \theta_{p1}$. The interaction of elastic vibrations $\tau \ll \tau_{p1}$ with such a process is one of the mechanisms of relaxation resonance [15]. If the sample contains the volume density $\rho_L$ of straight dislocation segments with length $L$, then their contribution to the dynamic modulus of elasticity and vibration decrement is described, without taking into account the statistical scatter of parameters, by formulas (9)-(12), in which

$$U = U_0 \simeq 2\varepsilon_k, \qquad \theta_0 = \theta_{p1}, \tag{22}$$

We assume that the discussed resonance $T_{ps} = 190$ K in the studied HEA is caused by the Seeger process, therefore, according to (13)

$$C_r\Delta_0 = \left(C_r\Delta_0\right)_L \approx 10^{-1}L^3\rho_L . \qquad (23)$$

Comparison of the values of parameters $U_0^s$ and $\theta_0^s$ obtained as a result of the analysis of experimental data (see Table 3) with the formulas of the theory (20)-(23) does not lead to contradictions and allows us to obtain estimates for the characteristics of dislocations responsible for resonance.

From (21) and (22) the next relations follow:

$$U_0 = 2c_t^2 m_k , \quad \frac{U_0}{\theta_{p1}} = \frac{4c_t b a_p}{\pi^2}\tau_{p1} , \quad U_0\theta_{p1} = \frac{8a_p^2}{\pi}\sqrt{\Gamma M} = \frac{8a_p^2}{\pi c_t}\Gamma = \frac{8c_t a_p^2}{\pi}M \qquad (24)$$

The necessary for further assessments parameters of the studied HEA have the following values:

$$b = a_0 = 2.54\cdot 10^{-10}\ \text{m} , \qquad a_p = \frac{\sqrt{3}}{2}a_0 \approx 0.87a_0 \approx 2.2\cdot 10^{-10}\ \text{m}\ [13];$$

$$\rho = 7.98\cdot 10^3\ \frac{\text{kg}}{\text{m}^3}\ [6], \quad G = \frac{E}{2(1+\nu)} = 0.94\cdot 10^{11}\ \text{Pa} , \quad c_t = \sqrt{\frac{G}{\rho}} \approx 3.4\cdot 10^3\ \frac{\text{m}}{\text{s}} \qquad (25)$$

here $G$ - shear modulus, Poisson's ratio $\nu \approx 0.22$ [7], $\rho$ - density. The resulting estimate $c_t = \sqrt{\frac{G}{\rho}} \approx 3.4\cdot 10^3\ \frac{\text{m}}{\text{s}}$ is in good agreement with the data [7].

Substituting into (21) -(24) the values of the activation energy $U_0 = U_0^s \approx 11.2\cdot 10^{-21}$ J, the period of attempts $\theta_0^s \simeq 4\cdot 10^{-11}$ s and $\left(C_r\Delta_0\right)_L \le 1\cdot 10^{-4}$ for resonance $T_{ps} = 190$ K leads to the following estimates for the parameters of the dislocation model under consideration:

$$m_k \approx 5\cdot 10^{-3}m_a \approx 5\cdot 10^{-28}\ \text{kg} , \quad \lambda_k \approx 40a_0 \approx 1\cdot 10^{-8}\ \text{m} , \quad \tau_{p1} \approx 3.6\cdot 10^6\ \text{Pa} \approx 4\cdot 10^{-5}G ,$$

$$M \approx 1.1\cdot 10^{-15}\ \frac{\text{kg}}{\text{m}} \approx 2.1\rho b^2 , \ \Gamma \approx 12.4\cdot 10^{-9}\ \frac{\text{J}}{\text{m}} \approx 2.1Gb^2 , \quad \rho_L L^3 \le 10^{-3} . \qquad (26)$$

Here the mass $m_k$ and width $\lambda_k$ of the kink are compared with the average mass $m_a$ of an atom of the studied alloy and the minimum interatomic distance $a_0$, and the Peierls critical stress $\tau_{p1}$ with the shear modulus $G$.

The good correspondence of the estimates for the parameters $\Gamma$ and $M$ with their estimates $\Gamma \approx Gb^2$ and $M \approx \rho b^2$ in the continuum theory of dislocations is a serious argument in favor of the adequacy of the proposed model of the relaxation process $T_{ps} = 190$ K to the experimentally recorded resonance properties.

The estimates we obtained are in good agreement with the estimates [5, 33] obtained from experiments on active deformation: low temperature values of the yield strength $\tau_{0.2}, \tau_2 \approx (2 \div 3) \cdot 10^8$ Pa $\approx 50 \cdot \tau_{p1}$, average area per local obstacle $S_0 = 1.7 \cdot 10^{-17}(b/w)^3$, m$^2$ ~ 261 $b^2$, $l = (S_0)^{1/2} = 16\, b = 4.1$ nm and $L = 2l \approx 10$ nm .

# 5. CONCLUSION

Analysis and physical interpretation based on modern dislocation theory of the results of a comprehensive experimental study [4-6, 33] of the processes of plastic deformation and acoustic relaxation in HEA $Al_{0.5}CoCrCuFeNi$ made it possible to establish:

- the most important types of dislocation defects in the lattice structure of the alloy;
- types of barriers that prevent the movement of dislocation lines (strings);
- adequate mechanisms of thermally activated movement of various elements of dislocation strings through barriers under conditions of moderate and deep cooling;
- quantitative estimates for the most important characteristics of dislocations and their interaction with barriers.

The authors are grateful to M.A. Tikhonovsky for the samples provided for research and E.D. Tabacnikova for valuable discussions.